# VALIDATION METHOD TO IMPROVE BEHAVIORAL FLOWS ON UML REQUIREMENTS ANALYSIS MODEL BY CROSS-CHECKING WITH STATE TRANSITION MODEL

Hikaru Morita [1] and Saeko Matsuura [1]

[1] Graduate School of Engineering and Science, Shibaura Institute of Technology, Minuma-ku 307, Saitama, Japan
`{mf19077, matsuura} @shibaura-it.ac.jp`

## ABSTRACT

*We propose a method to evaluate and improve the validity of required specifications by comparing models from different viewpoints. Inconsistencies are automatically extracted from the model in which the analyst defines the service procedure based on the initial requirement; thereafter, the analyst automatically compares it with a state transition model from the same initial requirement that has been created by an evaluator who is different from the analyst. The identified inconsistencies are reported to the analyst to enable the improvement of the required specifications. We develop a tool for extraction and comparison and then discuss its effectiveness by applying the method to a requirements specification example.*

## 1. INTRODUCTION

In recent years, a system that provides services is often complex, and it is linked with various hardware and other systems. To build a system that satisfies the final service goal, it is important to verify that the requirement specifications satisfy the goal after considering the characteristics of the system components in the requirements analysis phase. We studied model driven development, which defines requirements specifications using unified modeling language (UML) models based on use case analysis [1]. Further, we verified the inconsistencies between the models using the model-verification technique [2,3] and converted the investigated specifications into products at the design and implementation phases [4]. Similar to our approach [2], Tariq et al. [5], and Rafe et al. [6] transformed the activity diagram created in the requirements analysis stage into a finite state model suitable for model verification tools, i.e., a formal verification technique. Hence, exhaustive verification was performed to guarantee reachability and safety. However, although these verification methods can confirm the validity of the service procedure described, any excess or deficiency in the user request cannot be verified. To determine the excess or deficiency in a requirements specification, the requirements specification must be interpreted from multiple perspectives; in addition, requirements that are necessitated must be identified based on their differences.

Herein, we compare a requirements analysis model by defining the service procedure based on the consensus of multiple developers and the state transition model from the viewpoint of the state that the system should assume to perform the service. The purpose is to cross-validate the requirements analysis model and determine the excess or deficiency of the behavior.

The remainder of this paper is organized as follows. Section 2 describes the method to define a UML requirements analysis model and the role of the state transition model for evaluating the behavioral model. Section 3 describes the comparison method for the extracted and evaluation models. Section 4 discusses the effectiveness of our method using a case study. Finally, Section 5 discusses the conclusions and directions for future research.

## 2. REQUIREMENTS ANALYSIS MODEL AND EVALUATION MODEL

The service can be realized by linking the use cases provided by the system. In recent complex systems, the boundaries of each use case and the method to link the use cases at the early stage of development must be identified. We focused on exchanging information at each boundary in a system workflow and defined the cooperation and behavior of subsystems based on the procedure by the action flow using an activity diagram.

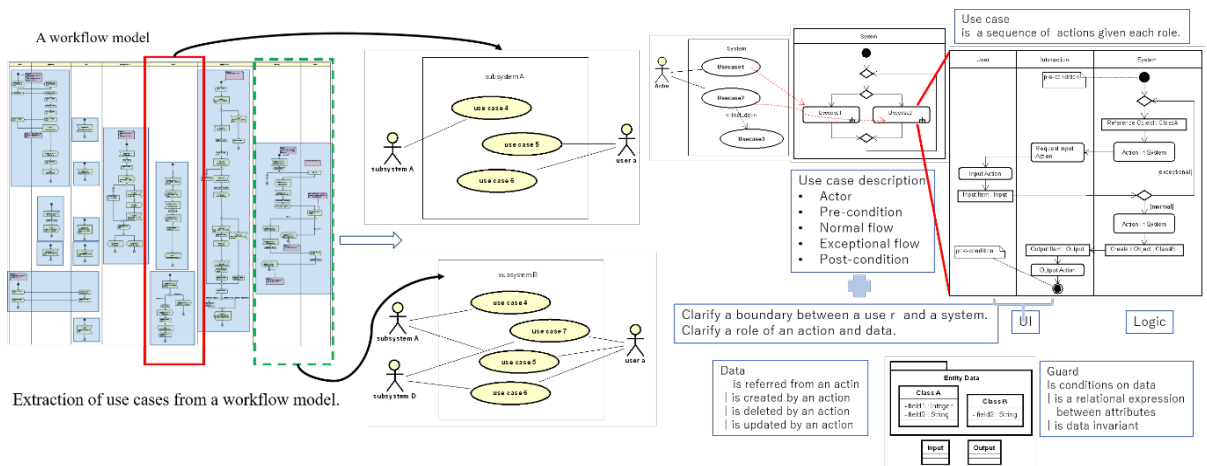

Figure 1: UML Requirements Analysis Model

A workflow is beneficial for realizing a service that utilizes the use cases of each subsystem. To achieve the system goals, the data exchange that is required to satisfy the service goals at each subsystem boundary must be clarified. Each partition in the activity diagram represents each subsystem and the users, and a set of use cases of each subsystem is described within the partition. Consequently, the cooperation between all users and subsystems is clarified, and the service procedure of the system is correctly defined.

Figure 1 shows the requirements analysis process after validating the workflow model. The use case diagram of each subsystem is described based on the partition of each subsystem of the workflow. At this point, the subsystem and the user who are exchanging data with the subsystem become actors in the use case diagram. Each use case not only defines the behavior by the activity diagram, but it also clarifies the behavior related to the data defined in the class diagram in Figure 1. We name this model *the UML requirements analysis model*. In such behavioral modeling, the procedure of the required function can be easily understood from the control structure. However, as the change in the state of the system due to the action is unclear, it is difficult to determine all the states that the system should assume.

Meanwhile, the state transition model defines the behavior by changing the state of the system by external or internal events. It is easy to understand whether the system requirements are comprehensively analyzed by specifying the state name; however, it is difficult to confirm the execution procedure of the system function and its relationship with other subsystems. Therefore, activity diagrams are suitable for defining the entire system workflow, which includes coordination between subsystems. On the contrary, the state transition model organizes the requirements that the subsystem should satisfy based on the state that the system should assume. Therefore, it can be a test case of requirements specification defined by the former; as such, we propose using it as the *evaluation model* of the requirements analysis model, as shown in Figure 2.

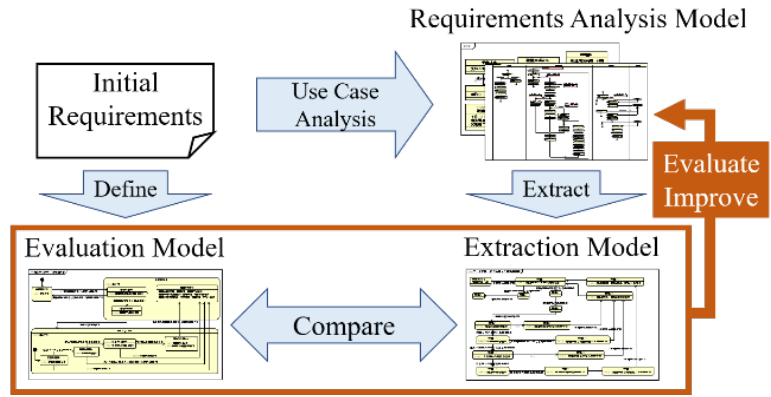

Figure 2: Validation Method by Comparing Extracted Model with Evaluation Model.

## 3. CONFIRMATION METHOD OF REQUIREMENTS SPECIFICATION S USING STATE TRANSITION MODEL

### 3.1. Extraction of State Transition Model from Workflow

The state was extracted from the requirements analysis model by focusing on the workflow control structure and actions. The extracted state transition model is named as the *extraction model*.

Table 1: Extraction Rules

| Activity model in workflow | State transition model | Activity model in workflow | State transition model |
|---|---|---|---|
| action0 → [guard0]/[guard1] → action1/action2 | state0 → [guard0]/[guard1] → state1/state2 | action0 → action1 → action2 | state0 → /action1 → state1 |
| action0 → <<signal receipt>> action1 → action2 | state0 → /action1 → state1 | action0 → <<process>> action1 → action2 | state0 → state1 do/action1 |
| action0 → time → action1 | state0 → time → state1 | Precondition: aaa → action0 | Precondition: aaa → state 0 |
| action0 → Update object A → objectA → action1 | state0 → state1 entry/Update object A | action0 → Postcondition: zzz | state 0 → Postcondition: zzz |

Table 1 shows the conversion rule from the requirements analysis model to the state transition model based on the following reason. First, the decision merge node is a transition control node and the transition differs depending on the branch condition; therefore, the separation of the states before and after that is identifiable.

Next, we focus on the action nodes. This is because the control structure divides the state by its guard, and it is assumed that the state changes by executing the action. Such actions comprise actions that receive data from other systems, actions that indicate the passage of time, actions that change the attributes of the system, and actions that are unique to the subsystems. Actions that receive data from other systems are described as *signal receiving nodes*, whereas actions that indicate the passage of time are described as *timers*. Signal-receiving nodes and timers are converted into events in the state machine diagram, *signal sending nodes*, and update actions, wherein the attributes of an object are converted into entry actions in a state; meanwhile, other actions are converted into actions on a transition arrow. Because the pre- and post-conditions directly represent the state of the system, they are used as the state of the state transition model. Figure 4 shows an example of the extraction model, which is expressed as a one-layer state transition model under the extraction rules. This extraction model is generated based on the rules in Table 1 from the activity diagram shown in Figure 3.

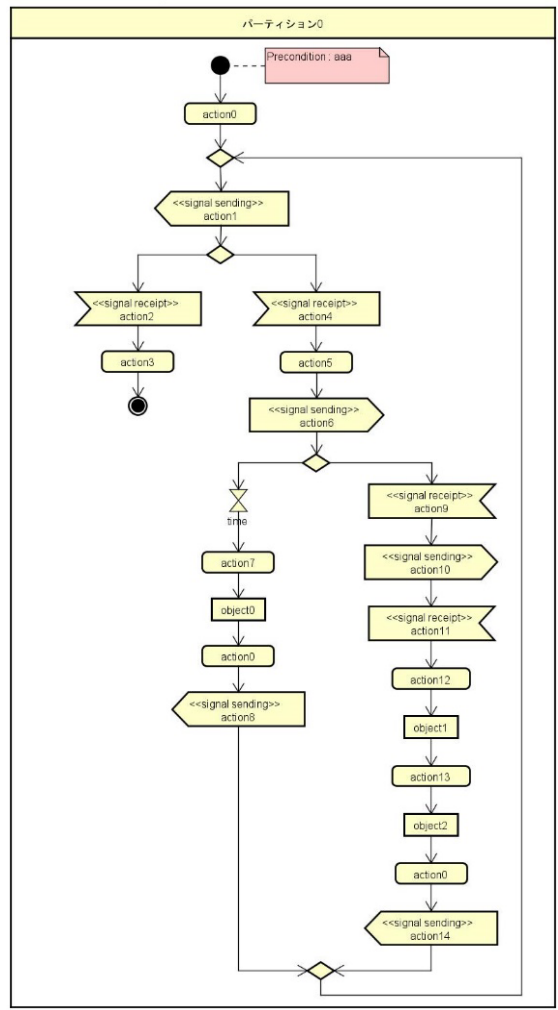

Figure 3: Example of Activity Model

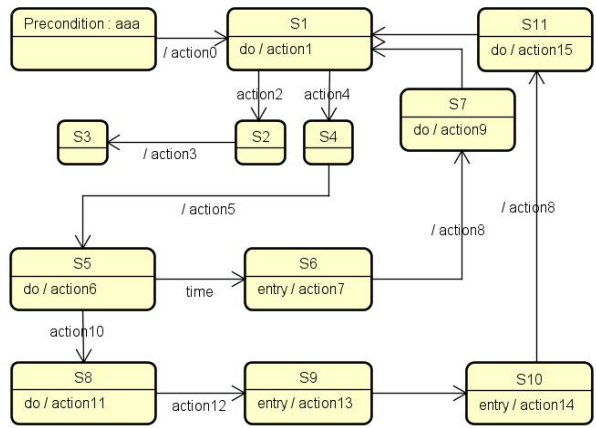

Figure 4: Example of Extraction Model

To confirm whether the extraction model includes the states to be assumed by the system indicated by the evaluation model, the states to be assumed by the system must be described appropriately based on the characteristics of the target system in the evaluation model. Because the workflow defines the cooperation scenario between subsystems, the state to be assumed by the subsystem should first be determined based on the state of each partner to be linked, as well as the type of work scenario to be executed within that state.

Consequently, the state of the evaluation model is defined by dividing it into layers for each viewpoint, as shown in Figure 5. In addition, if the words and phrases described in the evaluation model are freely described by the evaluator, it will be difficult to compare the contents of the state transition model. Therefore, we herein provide a list of actions of the defined workflow to unify words and phrases.

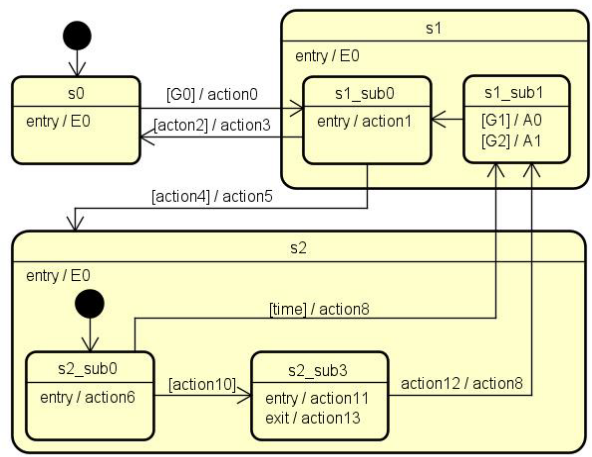

Figure 5: Example of Evaluation Model

### 3.2. Validation Process by Comparing Extraction and Evaluation Models

In the extraction model, the states are subdivided into one hierarchy based on the action and branch structure; therefore, each state is divided into the hierarchical states of the evaluation model based on a list of common actions prepared in advance. Table 2 provides a *judgment list* of behavior of the evaluation model shown in Figure 5.

Table 2: Judgment List of Behavior of Evaluation Model Shown in Figure 5

| s0 | s1 | s2 |
|---|---|---|
| **E0** | **E0** | **E0** |
| G0 | action1 | action6 |
| **action0** | action2 | time |
| | action3 | **action0** |
| | action4 | action9 |
| | action5 | action10 |
| | G1 | action12 |
| | A0 | action11 |
| | G2 | **action0** |
| | A1 | |

This list was generated for each hierarchy to be compared. All states, actions, and state transition contents, such as events, guards, and actions described in the state to be classified were acquired. Because the actions included in each state $s_i$ and the event guard actions described in the state transitions whose transition source $s_i$ occur in state $s_i$, they are classified into $s_i$. Duplicated labels listed in Table I are in bold font.

State transitions that transition between classification targets ($s_i$, i = 1, ... n) are beneficial for classifying hierarchies. We name these transitions "*external transitions.*"

Subsequently, the difference in interpretation between the two models is identified stepwise from the following viewpoints.

*Step 0: The name of the state in the extraction model is compared with the name of each state in the evaluation model.*

As shown by the rules listed in Table 1, the pre- and post-conditions in the workflow directly represent the states; therefore, they are compared with the state names of the extraction model obtained. Furthermore, by comparing between Figures 4 and 5, whether aaa matches any of the states can be determined.

*Step 1: The state of the extraction model is classified based on the operation described in the state.*

For each state in the extraction model, some behaviors included in it are verified if they can be classified into the state in the evaluation model using the judgment list. The category of the evaluation model that includes the state in which the behavior is described can be identified among the states of the extracted model. If the behavior is not described and the description element of the extraction model is not unique in the judgment list, then, the classification cannot be specified; consequently, it is set as unknown.

*Step 2: Classify by the event, guard, and action described in the state transition*

For an unknown state that cannot be classified by its own behavior, the state transition that exits from that state is acquired. The unknown state is classified by the event, guard, and action described in the acquired state transition using the judgment list. If no transition occurs, then, the description element is not unique in the judgment list, or no description element (we name it an unconditional transition) exists; consequently, the classification cannot be specified and hence, it is set as unknown. Table 3 lists the categories following the above mentioned three steps, and states divided by the first step are expressed in bold font.

Table 3: Classification by Steps 0,1, and 2

| s0 | s1 | s2 | unknown |
|---|---|---|---|
| **Precondition : aaa** | **s1** | **s5** | s3 |
| | s2 | **s8** | s6 |
| | s4 | **s9** | s7 |
| | | | s10 |
| | | | s11 |

*Step 3: Classify using an external transition based on the state before and after the state transition*

For the unknown state that has not yet been classified after the classification in the state transition, the previous state is acquired from the connected state transition. If the acquired state has already been classified into the state of the evaluation model, it is then classified into the same state. If it cannot be classified, the previous state is acquired from the state transition associated with the previous state, and the same judgment is performed. The acquisition of the previous state continues unless the behaviors on the connected state transition are the last in the list of behaviors of the abovementioned external transition. This is because the last behavior of the external transition divides the states in the same layer.

Table 4: Result of Step 3

| s0 | s1 | s2 | unknown |
|---|---|---|---|
| Precondition : aaa | s1 | s5 | |
| *s3* | s2 | s8 | |
| | s4 | s9 | |
| | **s7** | **s6** | |
| | **s11** | **s10** | |

Table 4 lists the results of the extraction model divided into the first layer of the evaluation model.

After this step, the extraction model is divided into the first layer in the evaluation model such that the number of states in this layer can be compared. Considering this difference, we can identify the following problems:

A) Among the states of the extraction model, a state exists that cannot be classified.

Therefore, the corresponding action of the workflow is not described in the evaluation model; hence, it may be in an unnecessary state that is derived from some excess behaviors. As the action flow causing the workflow defect is identified by tracing back the extraction rule, the flow can be rectified.

B) The extraction model does not contain a state that corresponds to the state of the evaluation model.

Therefore, the state considered in the evaluation model is not described as an action that can identify the state, such as a workflow signal reception action or an action for data. Some new actions must be added to identify the state of the workflow as the required state or flow may be missing.

C) The extracted model does not contain a state name that includes the state name of the evaluation model.

Therefore, the pre- and post-conditions may not be described in the workflow or the appropriate condition may not be defined. Hence, a note is added to the initial and final nodes of the workflow to add the pre- or post-condition.

After modifying the workflow or the evaluation model such that the number of states is equivalent, the number of state transitions of the modified extraction and evaluation models are compared.

The difference in the number of transitions is compared by generating a state transition table for both models. If the numbers differ, then the behaviors of the transitions are compared to identify the transitions that can be considered the same. The remaining transitions in the target state including some internal state transitions must be compared to determine whether their combinations are the same. In this case, the workflow transition condition may be ambiguous; therefore, some actions must be added. Moreover, if an additional transition exists, whether the behavior is necessary must be considered.

For all layers of the evaluation model, the previous steps are repeated to improve the workflow based on the observed difference. The improved extraction and evaluation models become a state transition model in which both the states and transitions are equal.

Figure 6 shows an extraction model after the abovementioned classification for all layers.

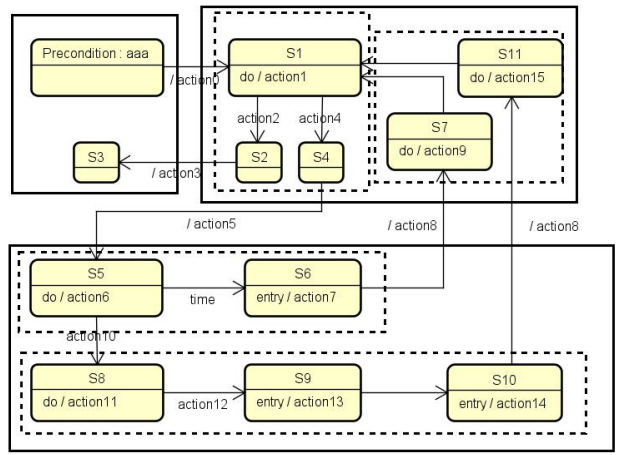

Figure 6: Classified Extraction Model

Finally, the correspondence of each behavior is verified, and the excess or deficiency of the behavior is determined. The difference between the two types of behavior models is noteworthy. In the behavior model based on the activity diagram, when the state changes by the change in the attribute of an object, it is defined by an action that changes the state. However, in the state transition model, it is typically expressed directly by the state name. Therefore, the meanings of the different expressions must be confirmed when comparing the descriptive elements in both models. In the correspondence between the evaluation model shown in Figure 5 and the classified extraction model shown in Figure 6, action7 and action14 in the extraction model can be interpreted as corresponding to G1 and G2 of the evaluation model, respectively.

By comparing the description elements, we can perform the following modifications:

- Rectify the behavioral expressions of action9 and action15 to A0 and A1, respectively.

- Because G0 is omitted in the extraction model, it is added to the workflow preconditions.

- Because behavior E0 is required in all states in the evaluation model, it should be a precondition in the workflow, or the behavior at the start of each state as an action should be added.

## 4. CASE STUDY AND DISCUSSION

This method was applied to the automatic baggage transport system, which is a PBL task in our department. This system is a parcel delivery system that links two autonomous vehicle-type robots with six subsystems: a relay station, headquarters, a reception desk, and the recipient's house. Figure 7 shows the circumstances and the relation between all subsystems.

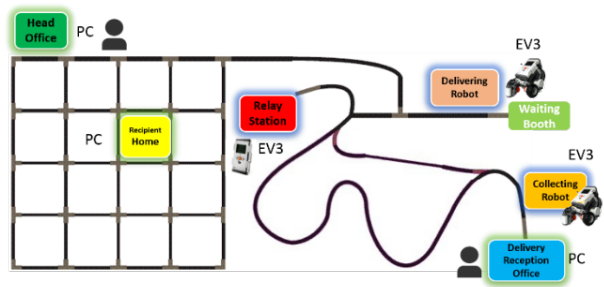

Figure 7: Circumstances and Relation Between Sub systems.

First, the workflow of this system is defined as a cooperation model of six subsystems. Figure 8 shows the extraction model generated from the workflow of the relay station based on the extraction rules. Figure 9 is an evaluation model that defines the requirements of the relay station as a state transition model. The orange box in Figure 8 shows the classification results based on the process outlined in Section 3 for the evaluation model. In this case, as a result of the state classification, the number of states in the first layer is equivalent to that in the evaluation model. However, as shown in Figures 7 and 8, the number of transitions between "waiting" to "working with delivery robot" differs. Comparing the state transitions inside both states, it was discovered that the difference in the transition branching is caused by the part surrounded by the dotted line in Figure 9.

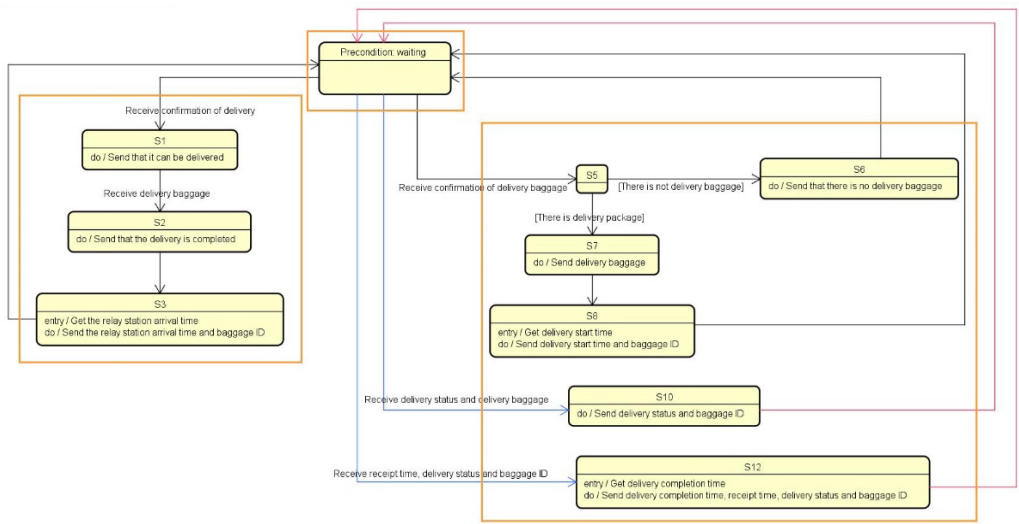

Figure 8: Extraction Model After Steps 0, 1, and 2

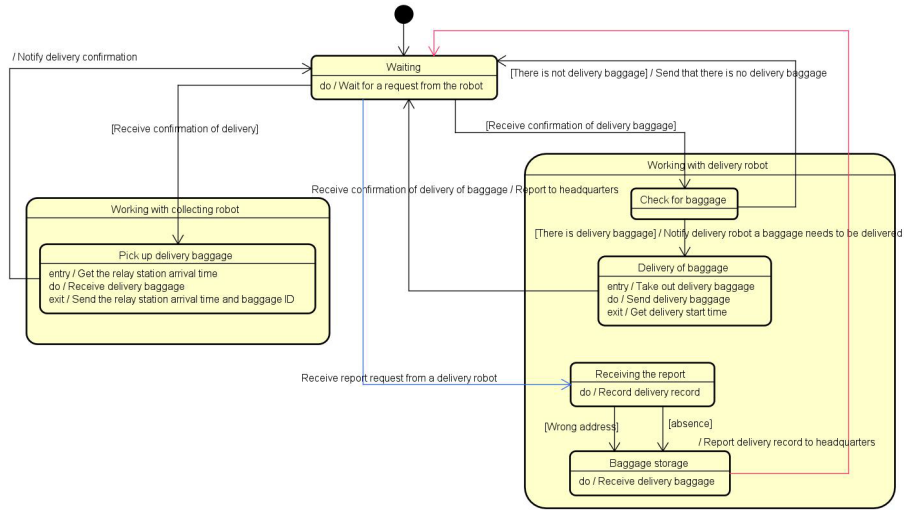

Figure 9: Evaluation Model

Comparing the state transitions inside both states, it is clear that the black transitions have the same meaning. However, because the remaining two transitions (see two sets of red and blue lines in Figure 8) have different transition branches depending on the part surrounded by the dotted line in Figure 9, a difference in meaning is generated.

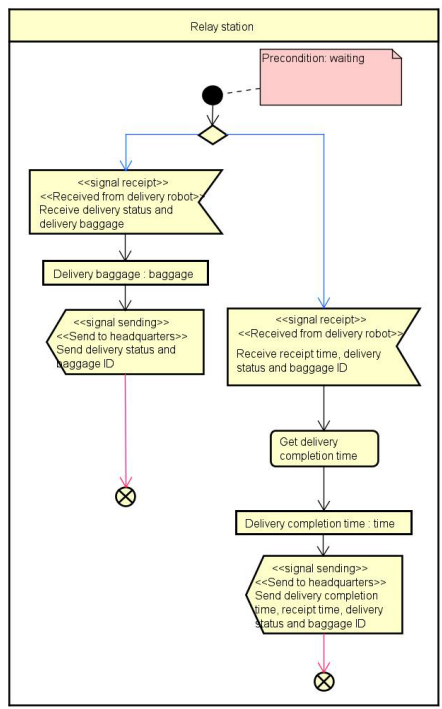

Figure 10: Corresponding Part of Workflow

Figure 10 shows a part of the workflow corresponding to the two blue lines in the extraction model. This can be automatically identified from states S10 and S12 based on the conversion rules.

Reviewing the workflow, it was discovered that the branching condition can be read from the branching processing flow, but the branching condition based on the delivery status value specified in the evaluation model was not specified in the workflow. Because the design model for the final program will be derived from the workflow, the processing procedure should be described in clear terms at this stage. Therefore, the workflow was modified as shown in Figure 11. Figure 12 shows the extraction model regenerated from the modified workflow.

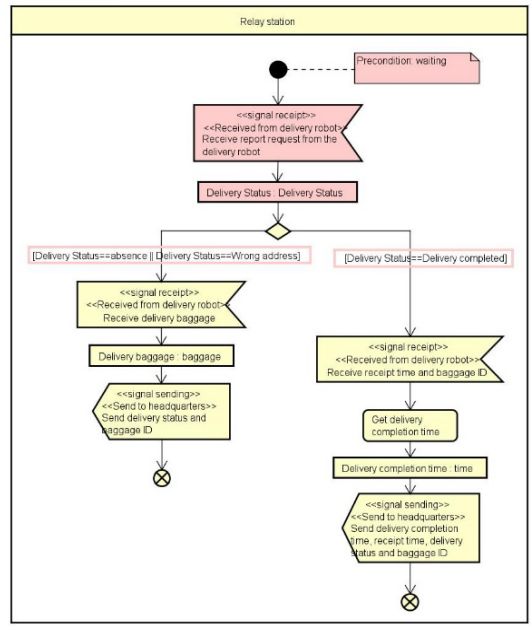

Figure 11: Improved Workflow

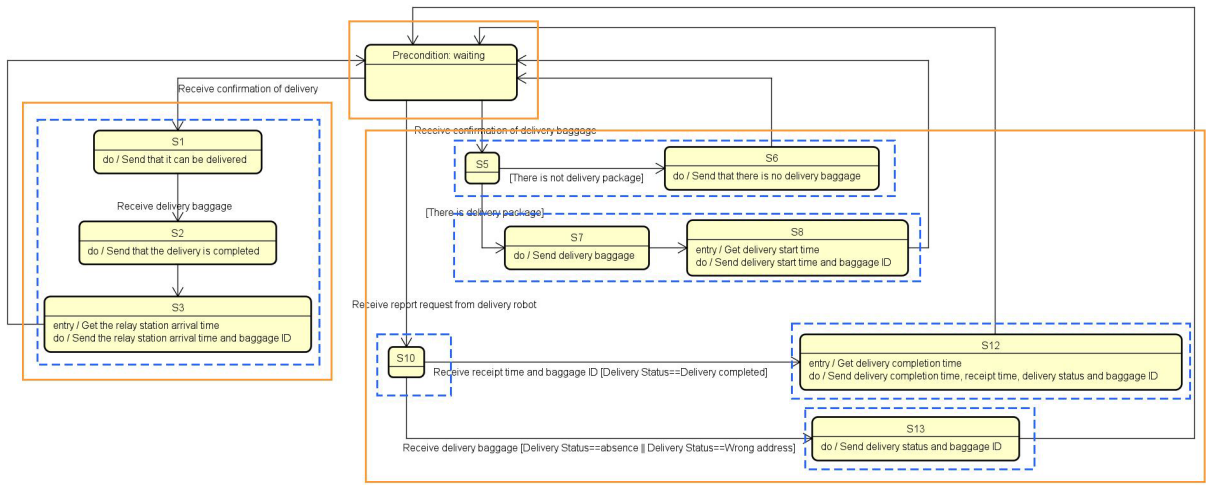

Figure 12: Regenerated Extraction Model

Next, we consider the transitions marked in red with states S10 and S12 as transition sources. The workflow defines the behavior of delivery completion, but the evaluation model does not include a behavior equivalent to delivery completion. As the initial requirements include the behavior at the time of delivery completion, it is clear that the evaluation model is insufficient. Figure 13 shows the modified evaluation model. The part surrounded by the red dotted line in Figure 13 denotes the added state and state transition.

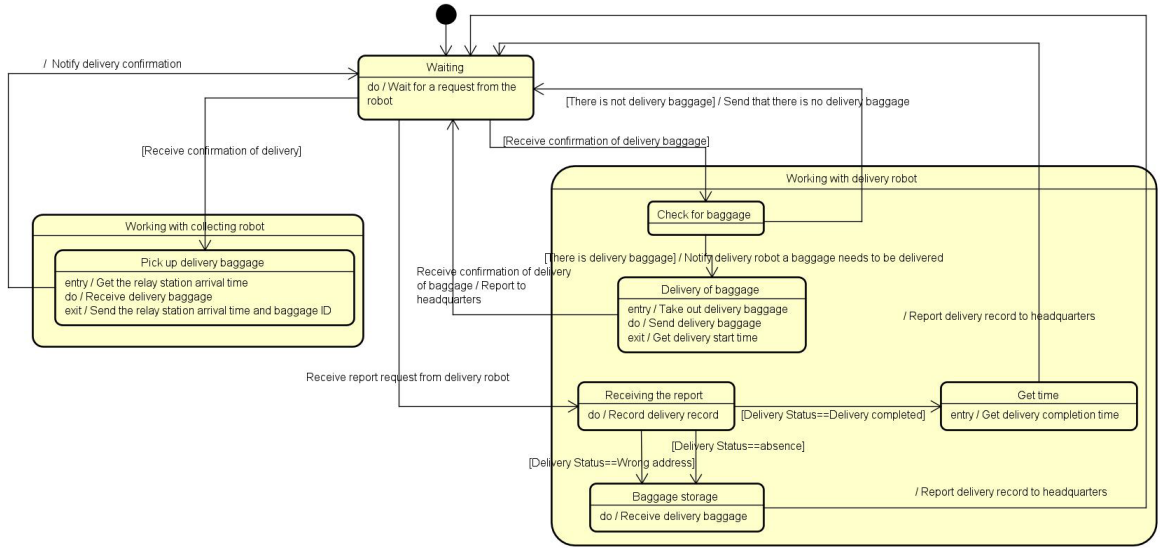

Figure 13: Improved Evaluation Model

After the first layer was modified, the extraction model was regenerated. It was observed that the number of states and transitions in the first layer was equivalent in both models.

Subsequently, the same classification was performed in the second layer. The results of steps 0, 1, and 2 are shown in Table 5, and the results of step 3 are shown in Table 6. Similarly, the number of states and transitions matched in the second layer.

Table 5: Classification by Steps 0, 1, and 2

| Check for baggage | Dlivery of baggage | Receiving the report | Baggage Storage | Get time | unknown |
|---|---|---|---|---|---|
| **S6** | **S7** | S10 | | **S12** | S11 |
| S5 | **S8** | | | | |

Table 6: Result of Step 3

| Check for baggage | Dlivery of baggage | Receiving the report | Baggage Storage | Get time | unknown |
|---|---|---|---|---|---|
| S6 | S7 | S10 | **S11** | S12 | |
| S5 | S8 | | | | |

As the workflow and evaluation model were improved by the difference for all layers of the evaluation model, the improved extraction and evaluation models became a state transition model, in which the states and transitions were equal. Finally, we compared the words of the event, guard, and action, which were the behavior of the transition or the state and determined the excess or deficiency of each behavior.

In reference to the "working with collecting robot" state shown in Figures 11 and 12, the four behaviors can be regarded as having the same content, but the behaviors circled in red showed different expressions. As the expression in the evaluation model was considered to be better, the expression of the action in the workflow was modified. Furthermore, it was discovered that the behavior order was different. The order in the evaluation model appeared to be more appropriate; therefore, the order of actions in the workflow was altered.

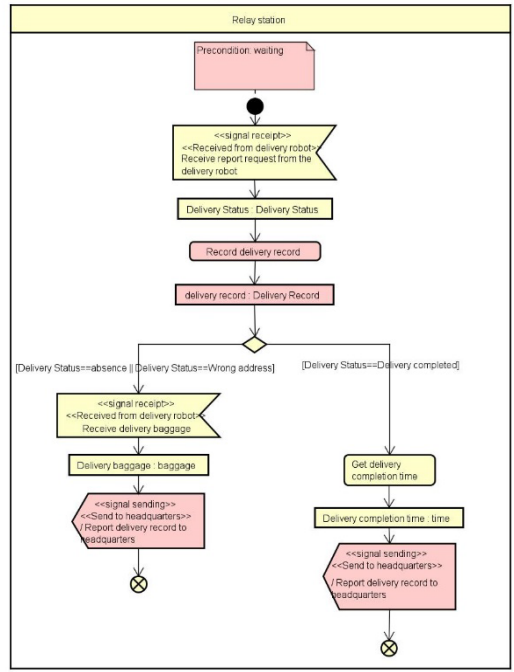

Figure 14: Modified Workflow

Next, we focus on the "working with delivery robot" state shown in Figures 11 and 12. The extraction model did not exhibit the behavior corresponding to the behavior circled in blue in the evaluation model. Because the behavior was insufficient, an action at the part corresponding to the relevant state of the workflow was added. Moreover, in the extraction model of Figure 12, the blue underlined behavior aims to update the "delivery record" such that the representation is modified more clearly. Figure 14 shows a workflow that reflects these modifications. The actions shown in pink in Figure 14 represent the added and modified actions.

## 5. CONCLUSIONS

To realize model-driven development in which behaviors are assigned to classes from the requirements analysis models and the final programs are generated from the designed class diagrams, the requirements analysis model must be of high quality. We proposed a validation method to improve the quality of requirements analysis models and developed a support tool.

The support tool was implemented as a plug-in in astah * Professional [7], which is a UML modeling tool. It offers the following two functions:

- A function to generate an extraction model from the workflow of the selected subsystem.

- A function to compare the extraction and evaluation models and present the information regarding the difference.

According to this information, the developer improves the behavior described in the workflow. However, a deficiency might be discovered in the interpretation of requirements in the evaluation model.

In the case study, we discovered differences in interpretation as well as the lack of descriptions such as guard conditions or some states that must be specified in the requirements analysis model. Because it is unclear whether the description of the evaluation model is always appropriate, it is possible that truly necessary and unnecessary requirements should be determined by comparing the two different view models and discuss the differences. We plan to apply this approach to various cases and verify its effectiveness.

**Authors**

Hikaru Morita

In 2020, enrolled in the master's program at the Department of Systems Science and Engineering, Graduate School of Science and Engineering, the same university. Currently engaged in research in the field of software engineering.

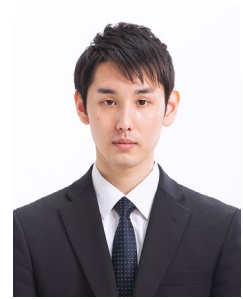

Saeko Matsuura

Shibaura Institute of Technology system Professor, Faculty of Science and Technology of Electronics and Information Systems Department in April 2013. Engaged in research on software development environments, design methodologies, and object-oriented development technologies.

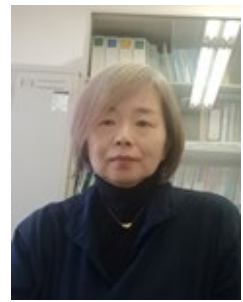